# Influence of light propagation on its polarization in free space

**Evelina Bibikova**[1,2,*] **and Nataliya Kundikova**[1,2]

[1]*Nonlinear Optics Laboratory, Institute of Electrophysics, Ural Branch of the Russian Academy of Sciences, 106, Amundsena Str., Ekaterinburg, 620016 Russia*
[2]*Department of Optics and Spectroscopy, South Ural State University, 76, Lenin Av., Chelyabinsk, 454080 Russia*

[*]*kundikovand@susu.ac.ru*

**Abstract:** We report the results of experimental observation of the influence of light propagation on its polarization under propagation in free space. These results experimentally prove the existence of the mutual influence of light polarization and propagation or spin-orbit interaction of a photon in free space. The value of the geometrical circular amplitude anisotropy $R$ appeared due to spin orbit interaction of a photon was measured and it turned out to be $R = \pm(0.60 \pm 0.08) \times 10^{-3}$.

**References and links**

1. S. M. Rytov, "On transition from wave to geometrical optics," Dokl. Akad. Nauk SSSR **18**, 263–266 (1938).
2. V. V. Vladimirskii, "The rotation of a polarization plane for curved light ray," Dokl. Akad. Nauk SSSR **21**, 222–225 (1941).
3. R. Y. Chiao and Y.-S. Wu, "Manifestations of berry's topological phase for the photon," Phys. Rev. Lett. **57**, 933–936 (1986).
4. A. Tomita and R. Y. Chiao, "Observation of berrys topological phase by use of an optical fiber," Phys. Rev. Lett. **57**, 937–940 (1986).
5. J. Picht, "Beitrag zur theorie der totalreflexion," Ann. Physik. **395(4)**, 433–496 (1929).
6. F. Goos and H. Hanchen, "Ein neuer und fundamentaler versuch zur totalreflexion," Ann. Physik. **436(7-8)**, 333–346 (1947).
7. F. Goos and H. Hanchen, "Neumessung des strahlversetzungseffektes bei totalreflexion," Ann. Physik. **440(3-5)**, 251–252 (1949).
8. F. I. Fedorov, "On the theory of total internal reflection," Dokl. Akad. Nauk SSSR **105(5)**, 465–469 (1955).
9. N. N. Kristoffel, "Total internal reflection and effects connected with it," Proc. Tartu Univ **42**, 94–111 (1956).
10. C. Imbert, "Calculation and experimental proof of the transverse shift induced by total internal reflection of a circularly polarized light beam," Phys. Rev. D. **5**, 787–796 (1972).
11. B. Y. Zel'dovich and V. S. Liberman, "Rotation of the plane of a meridional beam in a graded-index waveguide due to the circular nature of the polarization," Soviet Journal of Quantum Electronics **20**, 427 (1990).
12. A. V. Doogin, B. Y. Zel'dovich, N. D. Kundikova, and V. S. Liberman, "Effect of circular polarization on the propagation of light through an optical fiber," JETP Letters **53**, 197–199 (1991).
13. A. V. Dooghin, N. D. Kundikova, V. S. Liberman, and B. Y. Zel'dovich, "Optical magnus effect," Phys. Rev. A **45**, 8204–8208 (1992).
14. V. S. Liberman and B. Y. Zel'dovich, "Spin-orbit interaction of a photon in an inhomogeneous medium," Phys. Rev. A **46**, 5199–5207 (1992).
15. N. Kundikova, "Manifestation of spin-orbital interaction of a photon," Laser Physics **20**, 325–333 (2010).
16. K. Y. Bliokh and A. Aiello, "Goos-Hanchen and Imbert-Fedorov beam shifts: an overview," Journal of Optics **15**, 014001 (2013).
17. K. Y. Bliokh, A. Aiello, and M. Alonso, "Spin-orbit interactions of light in isotropic media," in *The Angular Momentum of Light,* D. L. Andrews and M. Babiker, eds. (Cambridge University, 2012), pp. 174245.

18. N. B. Baranova, A. Y. Savchenko, and B. Y. Zel'dovich, "Transverse shift of a focal spot due to switching of the sign of circular polarization," JETP Lett. **59**, 232–234 (1994).
19. B. Y. Zel'dovich, N. D. Kundikova, and L. F. Rogacheva, "Observed transverse shift of a focal spot upon a change in the sign of circular polarization," JETP Lett. **59**, 766–769 (1994).
20. N. D. Kundikova, F. V. Podgornov, L. F. Rogacheva, and B. Y. Zel'dovich, "Manifestation of spin-orbit interaction of a photon in a vacuum," Pure and Applied Optics: Journal of the European Optical Society Part A **4**, 179 (1995).
21. A. Y. Bekshaev, "Improved theory for the polarization-dependent transverse shift of a paraxial light beam in free space," Ukr. J. Phys. Opt. **12**, 10–18 (2011).
22. A. Aiello, N. Lindlein, C. Marquardt, and G. Leuchs, "Transverse angular momentum and geometric spin hall effect of light," Phys. Rev. Lett. **103**, 100401 (2009).
23. A. Y. Bekshaev, "Oblique section of a paraxial light beam: criteria for azimuthal energy flow and orbital angular momentum," Journal of Optics A: Pure and Applied Optics **11**, 094003 (2009).
24. K. Y. Bliokh, M. A. Alonso, E. A. Ostrovskaya, and A. Aiello, "Angular momenta and spin-orbit interaction of nonparaxial light in free space," Phys. Rev. A **82**, 063825 (2010).
25. M. Neugebauer, P. Banzer, T. Bauer, S. Orlov, N. Lindlein, A. Aiello, and G. Leuchs, "Geometric spin hall effect of light in tightly focused polarization-tailored light beams," Phys. Rev. A **89**, 013840 (2014).
26. Y. Zhao, J. S. Edgar, G. D. M. Jeffries, D. McGloin, and D. T. Chiu, "Spin-to-orbital angular momentum conversion in a strongly focused optical beam," Phys. Rev. Lett. **99**, 073901 (2007).
27. R. M. A. Azzam and N. M. Bashara, *Ellipsometry and polarized light* (North - Holland Pablishing Company, New York, 1977).
28. A. Gerrard and J. M. Burch, *Introduction to matrix methods in optics* (John Wiley and sons, New York, 1975).
29. I. Goltser, M. Darsht, N. Kundikova, and B. Zel'dovich, "An adjustable quarter-wave plate," Optics Communications **97**, 291 – 294 (1993).
30. M. Y. Darsht, I. V. Goltser, N. D. Kundikova, and B. Y. Zel'dovich, "Adjustable half-wave plate," Appl. Opt. **34**, 3658–3661 (1995).
31. S. N. Savenkov, V. V. Marienko, E. A. Oberemok, and O. Sydoruk, "Generalized matrix equivalence theorem for polarization theory," Phys. Rev. E **74**, 056607 (2006).
32. E. A. Bibikova, N. D. Kundikova, A. M. Popkova, and I. I. Popkov, "Effective parameters of composed polarization systems," Proceedings of SPIE **8011**, 80110U–80110U–6 (2011).
33. E. A. Bibikova and N. D. Kundikova, "Properties of an adjustable quarter-wave system under conditions of multiple beam interference." Applied optics **52**, 1852–1856 (2013).

## 1. Introduction

A polarized laser beam is known to carry a spin angular momentum that is associated with circular polarization and an orbital angular momentum that is associated with its way of propagation. The influence of a trajectory on light polarization [1, 2, 3, 4] and the influence of light polarization on a trajectory [5, 6, 7, 8, 9, 10] in an optically inhomogeneous medium were investigated independently during almost 60 years. For the first time the mentioned effects were considered as mutually inverse effects in 1990 [11] and the term spin-orbit interaction of a photon was introduced and used in 1991 [12] to explain experimentally observed Optical Magnus effect [12, 13]. After theoretical description of the Optical Magnus effect [14] the manifestation of the spin-orbit interaction of a photon in an inhomogeneous medium was considered in a rich array of a literature (see the most recent reviews [15, 16, 17]).

As for the spin-orbit interaction of a photon in an optically homogeneous medium (free space) the first prediction [18] and experimental observation [19] of the influence of polarization on light propagation were done in the following optical scheme. Let a circular polarized laser beam is fallen onto the bottom part of a lens ($y > 0$, Fig.1 ). The waist of the right circularly polarized ($\sigma = +1$) beam propagating in $z$ direction will be shifted in $+x$ direction. The waist of the left circular polarized ($\sigma = -1$) beam propagating in $z$ direction will be shifted in $-x$ direction. The change of the circularity sign of the beam leads to the waist shift. Really only $z$-polarized part of the field in the focal plane will be shifted upon the sign polarization change. The value of the shift of $z$-component observed in the scattering medium was about 1.5 $\mu$m [19, 20].

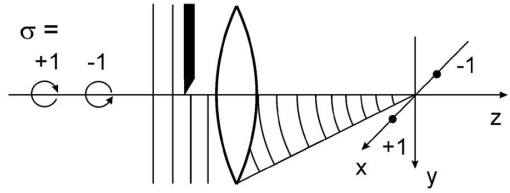

Fig. 1. Circularly polarized light propagation through the half of a lens (Ref. [20], Fig. 1). Here $\sigma = \pm 1$ is the sign of circular polarization, $\sigma = +1$ stands for right-handed circularly polarized light and $\sigma = -1$ stands for left-handed circularly polarized light.

The waist shift was predicted for the specific intensity distribution in the beam cross section [18]. The generalized theory which is applicable to a beam with an arbitrary intensity distribution in its cross section and which allows calculating the transverse waist shift depending on the sign of the circular polarization was proposed in [21]. The beam splitting into two beams with left-handed and right-handed circular polarization in a plane tilted with respect to the direction of propagation of the beam in free space was theoretically predicted in [22, 23], the geometric nature of a such splitting was shown and interpreted as geometric spin Hall effect of light [22]. Theoretical calculation of the beam waist intensity distributions was performed for a converging Bessel beam with nonzero topological charge. The obtained Bessel beam intensity distributions exhibit fine splitting of caustics [24].

Experimental investigations of geometric spin Hall effect of light were performed using a strongly convergent asymmetric circularly polarized beam. The beam cross-section was consisted of two parts with circular polarization of opposite signs. To observe the effect a nano-probing technique in combination with a reconstruction algorithm was used [25].

Gold nanoparticles were used for the experimental demonstration of spin-orbit interaction of a photon in a homogeneous isotropic medium. Gold nanoparticles were captured in the focal plane of a tightly focused circularly polarized "vortex" beam ($LG_{10}$). The particles rotation speed differed by 25% for circularly polarized light with different sign of circularity due to optical spin-to-orbital angular momentum conversion [26].

If the described effects [18, 19, 20, 21, 22, 23, 24, 25, 26] are the manifestation of the spin-orbit coupling of a photon in a homogeneous medium then the inverse effect, namely, the influence of propagation on polarization, should be observed.

In this paper we propose a scheme for observation of influence of light propagation on its polarization and report results of experimental observation of the change of linear polarization under the light trajectory change.

2. **An optical scheme for observation of the light propagation influence on its polarization**

Let us consider linearly polarized light propagation in $z$-direction through the bottom half ($y > 0$) of a lens (Fig.2d) ). Linear polarized light is superposition of two beams of equal intensity. The first one is a left-handed circularly polarized beam and the second one is a right-handed circularly polarized beam. The center of gravity of the $z$-component of the first beam will be shifted in $-x$ direction and the center of gravity of the $z$-component of the second beam will be shifted in $+x$ direction in the focal plane. If linearly polarized light propagates in $z$-direction through the upper half ($y < 0$) of a lens (Fig.2a) the center of gravity of the $z$-component of the first beam will be shifted in $+x$ direction and the center of gravity of the z-component of the second beam will be shifted in $-x$ direction in the focal plane [18, 19, 20].

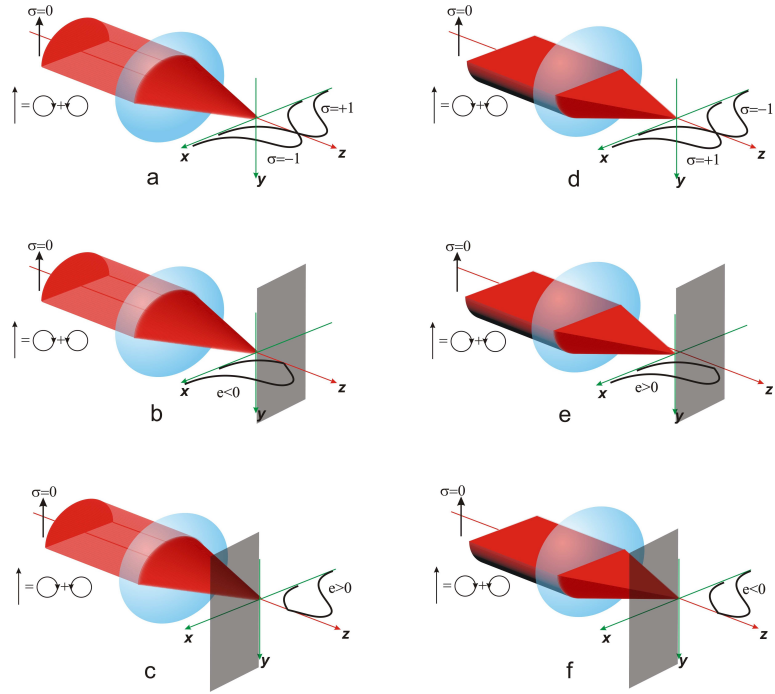

Fig. 2. An optical scheme for observation of the light propagation influence on its polarization. Here $e$ is ellipticity of elliptically polarized light, $-1 < e < +1$, $e < 0$ corresponds to left elliptically polarized light, $e > 0$ correspods to right elliptically polarized light and $e = 0$ corresponds to linearly polarized light.

Let us block the left half of the beam waist ($x > 0$) then the passed radiation will appear to be left-elliptically polarized ($e < 0$) if $y > 0$ (Fig.2f) or right-elliptically polarized ($e > 0$) if $y < 0$ (Fig.2c). Let us block the right half of the beam waist ($x < 0$) then the passed radiation will appear to be right-elliptically polarized ($e > 0$) if $y > 0$ (Fig.2e) or left-elliptically polarized ($e < 0$) if $y < 0$ (Fig.2b).

The described above transformation of linearly polarized light into elliptically polarized light is due to the different conditions for circularly polarized light propagation and can be considered as a result of influence of circular amplitude anisotropy [27, 31] with the value of the anisotropy $R$ of the different sign. The effect we are searching for can be named as geometrical circular amplitude anisotropy.

As it follows from the above speculation the ellipticity value is expected to be very small ($|e| \ll 1$). To carry out experimental investigation we should estimate the ellipticity value and choose the proper method of low ellipticities ($|e| \ll 1$) measurement.

To prove the existense of the predicted effect we should demonstrate the change of the sign of circular amplitude anisotropy $R$ under the change of the beam propagation.

### 3. Estimation of the value of the expected ellipticity

Let us estimate the value of the expected ellipticity of the light beam under investigation using the results presented in [18]. The intensity of the z-component of the converging beam is determined by the angular width $\theta_0$:

$$\theta_0 = (ka_0)^{-1} \tag{1}$$

and is about $\theta_0^2$. Here $k = 2\pi/\lambda$, $a_0 = \Delta r(\text{HW}e^{-1}\text{M})$ is the half-width of the beam focal waist by the criterion $e^{-1}$ of the intensity at a maximum. In order to increase the z-component intensity we should increase the angular width $\theta_0$, but according to Eq. (1) the angular width increasing leads to the beam focal waist width decreasing. Let a half of the focal waist is screened and the beam angular width is $2\theta \sim 0.16$ radian and the light wavelength $\lambda = 632.8$ nm. Then the angular factor is equal to $\theta^2 \sim 6 \cdot 10^{-3}$ and the radius of the beam focal waist $a_0 = \lambda/\pi\theta \sim 2.5$ $\mu$m.

Intensity distribution of z-components in the focal plane along $x$ axis $I_z = |E_z(x,0,0)|^2$ has the following form [18]:

$$I_z = |E_z(x,0,0)|^2 = \exp\left(\frac{-x^2}{a_0^2}\right) \frac{(1+\sigma x/a_0)^2}{(ka_0)^2}. \tag{2}$$

Here $\sigma = +1$ for right-handed circular polarized light and $\sigma = -1$ for left-handed circular polarized light.

If the beam is linearly polarized, it contains equal parts of the intensity of left- and right-handed circularly polarized components $I^R = I^L = I$. If the left half ($x > 0$) of the beam waist in the focal plane is blocked by the screen, the total intensity of the linear polarized beam and the intensity of the left-circular polarized light $I^L_{screen}$ will be halved and will lose the part of intensity $I_z$ corresponding to $z$-component:

$$I^L_{screen} = \frac{I}{2} - I_z. \tag{3}$$

The total intensity of right-circular polarized light after the focal plane $I^R_{screen}$ will be just reduced about two times:

$$I^R_{screen} = \frac{I}{2}. \tag{4}$$

Elliptically polarized light is a superposition of two beams with the right-handed and left-hand circularly polarization of different intensity. The relationship between the ellipticity $e$ of the beam and the intensity of its right-handed $I^R$ and left-handed $I^L$ circularly polarized components is the following :

$$e = \frac{\sqrt{I^R/I^L} - 1}{\sqrt{I^R/I^L} + 1} = \frac{1 - \sqrt{I^L/I^R}}{1 + \sqrt{I^L/I^R}}. \tag{5}$$

Using Eq.(3) and Eq.(4) we get:

$$e = \frac{\sqrt{1 - 2(I_z/I)} - 1}{\sqrt{1 - 2(I_z/I)} + 1}. \tag{6}$$

The maximum fraction of $z$-component intensity normalized to the intensity of the light beam at the maximum $I_z/2I$ is approximately 0.006. Insertion of 0.012 for $I_z/I$ into Eq. (6) gives the ellipticity of polarized light after the collimating lens. The acquired beam ellipticity $e$ of the initially linearly polarized beam will be $e \sim 6.0 \cdot 10^{-3}$.

To dealing with such low ellipticities $|e| \ll 1$ we should use proper method of low ellipticities measurement and be confident that the ellipticitity of a laser beam used in experimental set up is much less than $10^{-3}$.

### 4. Method of low ellipticity measurement

To measure the ellipticity $e$ of a laser beam a method of null ellipsometry is usually used [27]. The method allows to measure the values of ellipticity $e$ in the all range of ellipticities $-1 < e < +1$ due to the usage of a quarter wave plate as a compensator. The wide range of ellipticities leads to a moderate accuracy. To measure the beam ellipticities in the range $-6.0 \cdot 10^{-3} < e < 6.0 \cdot 10^{-3}$ with high accuracy we were using the following modified method of null ellipsometry.

Let us describe the polarization state conversion using Jones calculus [27, 28]. We will describe the polarization state of a fully polarized quasi-monochromatic light beam propagating along the *z*-direction by the Jones vector or the Maxwell vector:

$$E = \begin{pmatrix} E_x \\ E_y \end{pmatrix} \exp(i\omega t). \quad (7)$$

Here $i = \sqrt{-1}$, $E_x$ and $E_y$ are the complex components of the transverse electric vector along the *x*- and *y*- directions, respectively. The Jones vector for a linearly polarized beam with the angle $\beta$ between the direction of polarization and the *x*- direction has the form:

$$E_l = \begin{pmatrix} \cos\beta \\ \sin\beta \end{pmatrix}. \quad (8)$$

The Jones vector for an elliptically polarized beam with the angle of ellipticity $\varepsilon$ and the angle $\vartheta = 0$ between the major ellipse axis and the *x*-direction has the form:

$$E_e = \begin{pmatrix} \cos\varepsilon \\ i\sin\varepsilon \end{pmatrix}. \quad (9)$$

The angle of ellipticity $\varepsilon$ is connected with the light ellipticity *e* by the following way:

$$e = b/a = \tan\varepsilon. \quad (10)$$

Here *a* and *b* are the principal semi-axes of the polarization ellipse.

The optical system which transforms the polarization state linearly is described by a $2 \times 2$ Jones matrix with complex entries [27, 28].

The Jones matrix of a linear phase anisotropy, or a birefringent plate with mutually orthogonal slow and fast axes has the form

$$\hat{\mathbf{T}}^{LP}(\Gamma, \varphi) = \begin{pmatrix} \cos^2\varphi + \exp(i\Gamma)\sin^2\varphi & [1 - \exp(i\Gamma)]\cos\alpha\sin\varphi \\ [1 - \exp(i\Gamma)]\cos\varphi\sin\varphi & \sin^2\varphi + \exp(i\Gamma)\cos^2\varphi \end{pmatrix}, \quad (11)$$

where $\Gamma$ is phase shift between two orthogonal linear components of the electric vector, $\varphi$ is an azimuth of the linear phase anisotropy, $0 \leq \Gamma \leq 2\pi$ and $-\pi/2 \leq \varphi \leq \pi/2$.

If ellipticaly polarized light with the Maxwell vector described by Eq.(9) is transmitted through a phase plate described by Eq. (11) then we get the following Maxwell vector for the transmitted light beam:

$$\begin{pmatrix} (E_2)_x \\ (E_2)_y \end{pmatrix} = E_2 = \hat{\mathbf{T}}^{LP}(\Gamma, \varphi) E_1 \quad (12)$$

were

$$(E_2)_x = \cos(\Gamma/2)\cos\varepsilon - \sin(\Gamma/2)\sin 2\varphi\sin\varepsilon + i\sin(\Gamma/2)\cos 2\varphi\cos\varepsilon$$

and

$$(E_2)_y = \sin(\Gamma/2)\cos 2\varphi\sin\varepsilon + i\left[\cos(\Gamma/2)\sin\varepsilon + \sin(\Gamma/2)\sin 2\varphi\cos\varepsilon\right].$$

The light beam will be extinguished by an analyser if the beam transmitted through the compensator is linearly polarized and its ellipticity is equal to zero. Information about the ellipse of polarization can be extracted from the ratio [27, 28]:

$$\chi = \frac{(E_2)_y}{(E_2)_x}. \tag{13}$$

According to [27]

$$\chi = \frac{\tan\varphi + i\tan\varepsilon}{1 - i\tan\varphi\tan\varepsilon}. \tag{14}$$

If the transmitted light is linearly polarized than $\tan\varepsilon = 0$ and $\text{Im}\chi = 0$. Solving equation $\text{Im}\chi = 0$ we have:

$$\tan 2\varepsilon_1 = \sin 2\varphi \tan\Gamma. \tag{15}$$

Measuring the value of the compensator orientation $\varphi$ we get the ellipticity angle value $\varepsilon_1$:

$$\varepsilon_1 = \tan^{-1}\left(\frac{\sin 2\varphi \tan\Gamma}{2}\right). \tag{16}$$

Since $|\sin 2\varphi| \leq 1$, than $|2\varepsilon| \leq \tan^{-1}(\tan\Gamma)$, and the measured value of the ellipticity angle should be $|\varepsilon| \leq \Gamma/2$. Choosing the compensator with the retardation $\Gamma \approx 1° \approx 1.7 \times 10^{-2}$ and measuring the value of the compensator orientation $\varphi$ with the accuracy $0.1° \approx 1.7 \times 10^{-3}$ we can measure ellipticity $e \approx \varepsilon$ with the accuracy $\approx 10^{-5}$.

As it has been shown above the value of the expected ellipticity $e$ is approximately $6.0 \times 10^{-3}$. Commercially available He-Ne lasers can provide ellipticity approximately $0.5 \times 10^{-1}$. Usage of a Glan prism allows to decrease ellipticity up to $3.0 \times 10^{-3}$. To carry out investigation of the ellipticity change we should provide the ellipticity of the initial laser beam much less than $6.0 \times 10^{-3}$. It is possible to perform using a compensator with the value of retardation $\Gamma \approx 1°$.

Let us consider elliptically polarized light propagation through a phase plate with the retardation $\Gamma$ and the orientation angle $\varphi$. It easy to show that a phase plate with the angle of orientation $\varphi = \pi/4$ does not change the polarization ellipse orientation. Taking into account Eqs.(9,11) we get the following equation:

$$\begin{pmatrix} \cos\varepsilon_2 \\ i\sin\varepsilon_2 \end{pmatrix} = \hat{\mathbf{T}}^{\text{LP}}(\Gamma, \varphi = \pi/4)\begin{pmatrix} \cos\varepsilon_1 \\ i\sin\varepsilon_1 \end{pmatrix}. \tag{17}$$

Here $\varepsilon_1$ is the angle of ellipticity of an input beam and $\varepsilon_2$ is the angle of ellipticity of an output beam. Equation (17) gets the following form:

$$\begin{pmatrix} \cos\varepsilon_2 \\ i\sin\varepsilon_2 \end{pmatrix} = \begin{pmatrix} \cos(\Gamma/2)\cos\varepsilon_1 + \sin(\Gamma/2)\sin\varepsilon_1 \\ +i\left[\cos(\Gamma/2)\sin\varepsilon_1 - \sin(\Gamma/2)\cos\varepsilon_1\right] \end{pmatrix}. \quad (18)$$

Solving Eq.(18) we have:

$$\varepsilon_2 = (\varepsilon_1 - \Gamma/2). \quad (19)$$

According to Eq.(19) we can reduce the beam ellipticty just choosing the proper value of retardation $\Gamma$.

## 5. Ellipticity measurements

The experimental arrangement for investigation of the ellipticity change is shown in Fig.3. To form the converging light beam a He-Ne laser emitting at the wavelength $\lambda = 632.8$ nm was used. The laser beam power was $P = 35$ mW. The laser beam was expanded due to its own divergence during propagation. The distance between the laser and the first lens was $\sim 6$ m.

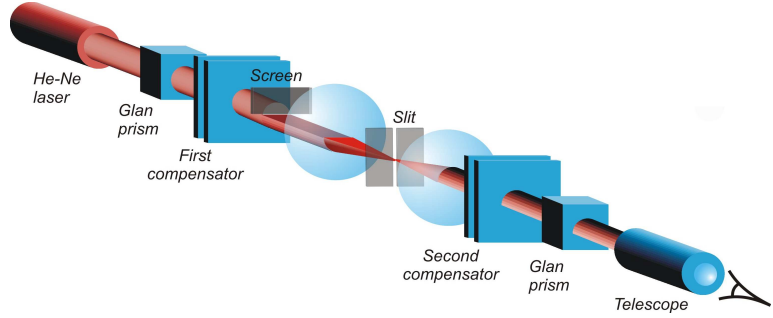

Fig. 3. Experimental setup for the ellipticity change measurements.

To decrease the initial laser beam ellipticity a Glan prism and a compensator with the retardation $\Gamma_c = 0.96°$ were used. The compensator was made from two mica plates according to the method described in [29, 30]. It was possible to improve the linear polarization of the light beam after passing through the polarizing Glan prism and the compensator up to $10^{-4}$.

The intensity distribution of the expanded beam in the plane of the focusing lens (Fig. 3) had a shape close to a Gaussian function. The beam diameter $d(\text{HW}e^{-2}\text{M})$ was approximately 6 mm by the criterion $e^{-2}$ of the intensity at the maximum. The expanded laser beam was focused by the first lens of the focal length $f = 3.8$ cm. The convergence angle $2\theta_{\text{conv}}$ of the generated beam was $2\theta_{\text{conv}} = 0.16$ radian.

It should be stressed that to determine the beam polarization state it was necessary to collimate the light beam under investigation. To collimate the light beam the second lens was placed after the focusing lens. The focal length of the collimating lens was equal to the focal length of the focusing lens. Alignment of the collimating lens was done using a telescope focused for infinity.

A screen was placed just ahead of the focusing lens. That screen was used to screen the upper or the bottom part of the beam to generate an asymmetric beam. The knife of a spectrometer slit was used to screen the right or the left part of the beam focal waist. The slit adjustment micrometer was used to place the slit knife edge into the center of the beam focal waist. The edge of the slit knife was set vertically. The knife slit was mounted on a translation stage to provide the movement along the beam.

To observe the effect the edge of the knife should be placed precisely in the focal plane, namely, in the middle of the beam waist. The beam waist diameter $d_\text{w}(\text{HW}e^{-2}\text{M})$ turns out to be $d_\text{w} \sim 15$ $\mu$m and the waist length was approximately 2 mm. Those values define the requirement to the adjusting optical system precision.

To measure the ellipticity change under the light propagatiog along the different way we used the described above method. A wave-plate with a retardation of $\Gamma_\text{c} = 1°$ was used as a compensator. The compensator was made from two mica plates according to the method described in [29, 30]. The compensator retardation was measured beforehand. The second Glan prism and a telescope were installed after the second compensator. To measure the light ellipticity the second compensator and the second Glan prism were rotated till the full light extinguishing. The light extinguishing was observed using the telescope. The value of the resulting angle of the compensator orientation provided the value of the ellipticity according to Eq.(16).

We measured the beam ellipticity before the first lens, after the beam propagation through the first lens without the screen and the slit knife and with different position of the screen and the slit knife. It should be stressed, that we observed the ellipticity change only, the orientation of the polarization ellipse was unchanged. As was mentioned above the light ellipticity before the lens was $10^{-4}$. The light ellipticity measured after the lens was $1.4 \times 10^{-3}$. The measured values of the ellipticities aquared by the light beam under propagation along the different trajectories are shown in Tabel 1. One can see the change of the ellipticity under the trajectory change, but the sign of the ellipticity was not changed.

To explain the observed results and estimate the value of the circular let us describe our system using generalized matrix equivalence theorem for polarization theory [31]. According to [31] any combination of elements with linear and circular phase and linear and circular amplitude anisotropy is equivalent to an optical system containing only one element of each kind in the order:

$$\hat{\mathbf{T}}^\text{Gen} = \hat{\mathbf{T}}^\text{CP} \times \hat{\mathbf{T}}^\text{LP} \times \hat{\mathbf{T}}^\text{CA} \times \hat{\mathbf{T}}^\text{LA}. \qquad (20)$$

Table 1. Beam ellipticity under the different ways of propagation.

|  | Light propagets through the lens upper half | Light propagets through the lens bottom half |
|---|---|---|
| Light propagets through the left part of the focal sport | $e^{UL} =$ $1.10 \times 10^{-3}$ | $e^{BL} =$ $2.17 \times 10^{-3}$ |
| Light propagets through the right part of the focal sport | $e^{UR} =$ $2.50 \times 10^{-3}$ | $e^{BR} =$ $1.20 \times 10^{-3}$ |

Here $\hat{\mathbf{T}}^{LA}$ is the Jones matrix of linear amplitude anisotropy:

$$\hat{\mathbf{T}}^{LA}(P,\theta) = \begin{pmatrix} \cos^2\theta + P\sin^2\theta & (1-P)\cos\theta\sin\theta \\ (1-P)\cos\theta\sin\theta & \sin^2\theta + P\cos^2\theta \end{pmatrix}, \quad (21)$$

where $P$ is a value of linear amplitude anisotropy (relative absorption of two linear orthogonal components of the electric vector) and $\theta$ is an azimuth of linear amplitude anisotropy, $0 \leq P \leq 1$ and $-\pi/2 \leq \theta \leq \pi/2$. $\hat{\mathbf{T}}^{CA}$ is the Jones matrix of circular amplitude anisotropy:

$$\hat{\mathbf{T}}^{CA}(R) = \begin{pmatrix} 1 & -iR \\ iR & 1 \end{pmatrix}, \quad (22)$$

where $R$ is a value of the circular amplitude anisotropy (relative absorption of two orthogonal circular components of the electric vector) and $-1 \leq R \leq 1$. $\hat{\mathbf{T}}^{LP}$ is the Jones matrix of linear phase anisotropy described by Eq.(11). $\hat{\mathbf{T}}^{CP}$ is the Jones matrix of circular phase anisotropy:

$$\hat{\mathbf{T}}^{CP}(\phi) = \begin{pmatrix} \cos\phi & \sin\phi \\ -\sin\phi & \cos\phi \end{pmatrix}, \quad (23)$$

where $\phi$ is a phase shift introduced for two orthogonal circular components of the electric vector or optical activity, $0 \leq \phi \leq 2\pi$.

As it was mentioned above, the ellipse orientation was not changed under the light propagation through the experimental scheme. It is easy to show that the circular phase anisotropy does not change ellipse orientation if $\phi = 0$. The linear phase anisotropy does not change ellipse orientation if $\varphi = \pi/4$. The linear amplitude anisotropy of the system under consideration is due to multibeam interference and $P = 1 - \delta P$ where $\delta P \ll 1$ [32, 33] and the linear amplitude anisotropy does not change ellipse orientation if $\theta = 0$. As it follows from the Table 1 we are dealing with small ellipticities and we can assume that $\sin\varepsilon \approx$

$\varepsilon \approx e \ll 1$ and $\sin\Gamma \approx \Gamma \ll 1$, $\cos\Gamma \approx 1$. Under these assumptions the general matrix $\hat{\mathbf{T}}^{Gen}(P,\theta,R,\Gamma,\varphi,\phi)$ has the following form:

$$\hat{\mathbf{T}}^{Gen}\left(\delta P, \theta=0, R, \Gamma, \varphi=\pi/4, \phi=0\right) = \\ = \hat{\mathbf{T}}^{CP}(\phi=0) \times \hat{\mathbf{T}}^{LP}(\Gamma, \varphi=\pi/4) \times \hat{\mathbf{T}}^{CA}(R) \times \hat{\mathbf{T}}^{LA}(\delta P, \theta=0). \quad (24)$$

If we take into account only members of the first order we get:

$$\hat{\mathbf{T}}^{Gen}\left(\delta P, \theta=0, R, \Gamma, \varphi=\pi/4, \phi=0\right) = \\ = \begin{pmatrix} 1 & i(\Gamma/2 - R) \\ i(\Gamma/2 + R) & 1 - \delta P \end{pmatrix}. \quad (25)$$

The equation of the ellipticity transformation can be written in the following form:

$$\begin{pmatrix} 1 \\ ie_{out} \end{pmatrix} = \begin{pmatrix} 1 & i(\Gamma/2 - R) \\ i(\Gamma/2 + R) & 1 - \delta P \end{pmatrix} \begin{pmatrix} 1 \\ ie_{in} \end{pmatrix}. \quad (26)$$

Solving Eq. (26) and taking into account only members of the first order we get:

$$e_{out} = \Gamma/2 + R + e_{in} \quad (27)$$

or

$$e_{out} = \Gamma/2 \pm |R| + e_{in} \quad (28)$$

Let us assume that the values $e^{UL}$ and $e^{BR}$ are due to the negative value of $R$ and the values $e^{UR}$ and $e^{BL}$ are due to the positive value of $R$. It is allows as to write down four systems of equation:

$$\begin{aligned} e^I &= \Gamma/2 + |R| + e_{in}, \\ e^D &= \Gamma/2 - |R| + e_{in}, \end{aligned} \quad (29)$$

were $e^I$ stands for $e^{UR}$, $e^{BL}$ and $e^D$ stands for $e^{UL}$, $e^{BR}$.

Solving these systems we get:

$$\Gamma = e^D + e^I - 2e_{in} \quad (30)$$

and

$$|R| = \frac{(e^I - e^D)}{2}. \quad (31)$$

Using the values of $e^{UR}$, $e^{BL}$ $e^{UL}$, $e^{BR}$ from the Table 1 we get four values for $\Gamma$ and four values for $|R|$, namely, $\Gamma_1 = 3.07 \times 10^{-3}$, $\Gamma_2 = 3.50 \times 10^{-3}$, $\Gamma_3 = 3.40 \times 10^{-3}$, $\Gamma_4 = 3.17 \times 10^{-3}$, $|R|_1 = 0.54 \times 10^{-3}$, $|R|_2 = 0.65 \times 10^{-3}$, $|R|_3 = 0.70 \times 10^{-3}$, $|R|_4 = 0.49 \times 10^{-3}$.

Averaging the values of $\Gamma_i$ and $|R|_i$ we get $\langle\Gamma\rangle = (3.29 \pm 0.17) \times 10^{-3}$ and $\langle|R|\rangle = (0.60 \pm 0.08) \times 10^{-3}$.

## 6. Conclusion

We were successful to observe experimentally the influence of propagation on polarization in free space and so experimentally proved the mutual influence of polarization and propagation or spin-orbit interaction of a photon in free space. The value of geometrical circular amplitude anisotropy $R$ appeared due to spin orbit interaction of a photon turned out to be $R = \pm(0.60 \pm 0.08) \times 10^{-3}$.

## Acknowledgment

This work has been partially performed in the framework of programs of fundamental research of the Physics Division of RAS Fundamental problems of photonics and physics of new optical materials (12-T-2-1003).